\documentclass[fleqn,11pt]{wlscirep}
\usepackage[utf8]{inputenc}
\usepackage[T1]{fontenc}
\usepackage{hyperref}

\usepackage[final]{pdfpages}
\makeatletter
\AtBeginDocument{\let\LS@rot\@undefined}
\makeatother
\title{Age and market capitalization drive large price variations of cryptocurrencies}

\author[1,+]{Arthur A. B. Pessa} 
\author[2,3,4,5,6,*]{Matja{\v z} Perc} 
\author[1,$\dagger$]{Haroldo V. Ribeiro} 

\affil[1]{Departamento de F\'isica, Universidade Estadual de Maring\'a - Maring\'a, PR 87020-900, Brazil}

\affil[2]{Faculty of Natural Sciences and Mathematics, University of Maribor, Koro{\v s}ka cesta 160, 2000 Maribor, Slovenia}
\affil[3]{Department of Medical Research, China Medical University Hospital, China Medical University, Taichung, Taiwan}
\affil[4]{Alma Mater Europaea, Slovenska ulica 17, 2000 Maribor, Slovenia}
\affil[5]{Complexity Science Hub Vienna, Josefst{\"a}dterstra{\ss}e 39, 1080 Vienna, Austria}
\affil[6]{Department of Physics, Kyung Hee University, 26 Kyungheedae-ro, Dongdaemun-gu, Seoul, Republic of Korea}

\affil[+]{email: arthur\_pessa@hotmail.com}
\affil[*]{email: matjaz.perc@gmail.com}
\affil[$\dagger$]{email: hvr@dfi.uem.br}

\begin{abstract}
Cryptocurrencies are considered the latest innovation in finance with considerable impact across social, technological, and economic dimensions. This new class of financial assets has also motivated a myriad of scientific investigations focused on understanding their statistical properties, such as the distribution of price returns. However, research so far has only considered Bitcoin or at most a few cryptocurrencies, whilst ignoring that price returns might depend on cryptocurrency age or be influenced by market capitalization. Here, we therefore present a comprehensive investigation of large price variations for more than seven thousand digital currencies and explore whether price returns change with the coming-of-age and growth of the cryptocurrency market. We find that tail distributions of price returns follow power-law functions over the entire history of the considered cryptocurrency portfolio, with typical exponents implying the absence of characteristic scales for price variations in about half of them. Moreover, these tail distributions are asymmetric as positive returns more often display smaller exponents, indicating that large positive price variations are more likely than negative ones. Our results further reveal that changes in the tail exponents are very often simultaneously related to cryptocurrency age and market capitalization or only to age, with only a minority of cryptoassets being affected just by market capitalization or neither of the two quantities. Lastly, we find that the trends in power-law exponents usually point to mixed directions, and that large price variations are likely to become less frequent only in about 28\% of the cryptocurrencies as they age and grow in market capitalization.
\end{abstract}
\begin{document}
\rfoot{\small\sffamily\bfseries\thepage/16}%

\flushbottom
\maketitle
\thispagestyle{empty}

\section*{Introduction}~\label{sec:intro}

Since the creation of Bitcoin in 2008~\cite{nakamoto2008bitcoin}, various different cryptoassets have been developed and are now considered to be at the cutting edge of innovation in finance~\cite{burniske2017cryptoassets}. These digital financial assets are vastly diverse in design characteristics and intended purposes, ranging from peer-to-peer networks with underlying cash-like digital currencies (\textit{e.g.} Bitcoin) to general-purpose blockchains transacting in commodity-like digital assets (\textit{e.g.} Ethereum), and even to cryptoassets that intend to replicate the price of conventional assets such as the US dollar or gold (\textit{e.g.} Tether and Tether Gold)~\cite{eichengreen2019commodity, bullman2019search}. With more than nine thousand cryptoassets as of 2022~\cite{coinmarketcap}, the total market value of cryptocurrencies has grown massively to a staggering \$2 trillion peak in 2021~\cite{cnbc}. Despite long-standing debates over the intrinsic value and legality of cryptoassets~\cite{kiviat2015beyond}, or perhaps even precisely due to such controversies, it is undeniable that cryptocurrencies are increasingly attracting the attention of academics, investors, and central banks, around the world~\cite{lo2014bitcoin, auer2022central}.

Moreover, these digital assets have been at the forefront of sizable financial gains and losses in recent years~\cite{winklevoss2020case, willing2022ethereum}, they have been recognized as the main drivers of the brand-new phenomena of cryptoart and NFTs~\cite{vasan2022quantifying, nadini2021mapping}, but also as facilitators of illegal activities, such as money laundering and dark trade~\cite{christin2013traveling, elbahrawy2020collective, branwen2022darknet}. Financial research dedicated to cryptoassets, on the other hand, has been mostly concerned with the extension of fairly traditional analyses~\cite{corbet2019cryptocurrencies}, including market efficiency~\cite{urquhart2016inefficiency, nadarajah2017inefficiency, wei2018liquidity, bariviera2017stylized, sigaki2019clustering, wu2020long}, distribution of price returns~\cite{osterrieder2017statistical, osterrieder2017bitcoin, begusic2018scaling}, and volatility~\cite{sapuric2014bitcoin, zhang2018stylized}. Researchers have also probed the hedging and safe haven capabilities of cryptoassets when combined with a portfolio of stocks~\cite{feng2018cryptocurrencies, conlon2020cryptocurrencies}, their behavior in the scenario of generalized market turmoil caused by the COVID-19 pandemic~\cite{corbet2020contagion, conlon2020cryptocurrencies}, and the formation of price bubbles~\cite{fry2016negative, garcia2014digital}.

Among these subjects, the distribution of price returns, especially of large price variations, is considered fundamental for evaluating this new market's intrinsic risks and modeling its dynamics~\cite{phillip2018anewlook, gronwald2014economics}. Earlier analyses have consistently found price returns to follow heavy-tail distributions. Chu \textit{et al.}~\cite{chu2015statistical} have adjusted a large number of probability distributions to the log-returns of daily prices of Bitcoin from 2011 to 2014, finding the generalized hyperbolic distribution (a heavy-tailed distribution) to be the best description of the data. Using daily prices of eight cryptoassets (Bitcoin, Dash, Ethereum, Litecoin, NEM, Stellar, Monero, and Ripple) and the Jarque-Bera test, Zhang \textit{et al.}~\cite{zhang2018stylized} have rejected the normality of their log-returns. Similarly, Osterrieder \textit{et al.}~\cite{osterrieder2017bitcoin} have also found the normal distribution to be incompatible with price returns of six cryptocurrencies (Bitcoin, Dash, Litecoin, MaidSafeCoin, Monero, and Ripple) over a three-year period (2014-2016). Feng~\textit{et al.}~\cite{feng2018cryptocurrencies} have fitted a generalized Pareto distribution to two years of daily price returns of seven cryptocurrencies (Bitcoin, Dash, Ethereum, Litecoin, Monero, NEM, and Ripple) and observed an asymmetry between the left and right tails. Finally, using high-resolution data obtained from exchanges and referring to different semesters between 2010 and 2018, Begu{\v s}i{\'c} and coworkers~\cite{begusic2018scaling} have found power laws to be plausible fits to the empirical distributions of large price variations of Bitcoin. This latter approach belongs to the field of econophysics~\cite{mantegna1999introduction}, and has established intriguing regularities in the distributions of log-returns of traditional financial assets such as a power-law distribution [$p(r) \sim r^{-\alpha}$] with typical exponents $\alpha \sim 4$~\cite{gopikrishnan1998inverse, mantegna1999introduction, gabaix2003theory}. 

The above short review of pertinent past research shows that, while the return distributions of cryproassets have attracted considerable interest, most previous works have investigated these distributions using data spanning only a few years of price history from small sets of cryptocurrencies (usually Bitcoin and a handful of the biggest cryptocurrencies by market capitalization). Moreover, past research has not established whether return distributions change over time and whether they are dependent on market capitalization. The main goal of this work is therefore to fill these gaps by presenting a dynamic analysis of the return distributions of more than seven thousand cryptocurrencies. 

Our results show that the vast majority of cryptocurrencies have return distributions with tails well described by power-law functions over their entire history. The typical values of the power-law exponents characterizing these distributions are smaller than those observed in traditional assets, showing that cryptoassets are more susceptible to large price variations, with about half of them not presenting a characteristic scale for price returns. Moreover, these tail exponents reveal an asymmetry in large price movements often characterized by smaller exponents for positive returns; that is, large positive price variations are expected to occur more frequently than negative ones in most cryptoassets, but this asymmetry is minimal for a few classes of cryptoassets such as stablecoins. Our research further demonstrates that changes in the tail exponents are often associated with the age of cryptocurrencies and their market capitalization, or only with age, with only a minority of cryptoassets affected by market capitalization alone or entirely unaffected by these two quantities. For digital assets affected by age or market capitalization, we find power-law exponents to have mixed directions, with about 28\% of all cryptocurrencies, and 37\% of the current top 200 cryptocurrencies, becoming less likely to exhibit large price variations as they age and grow in market capitalization. This in turn indicates that large price variations are expected to become less likely only for a small part of the cryptocurrency market.

\section*{Results}~\label{sec:results}

\begin{figure*}[!ht]
\centering
\includegraphics[width=1\linewidth]{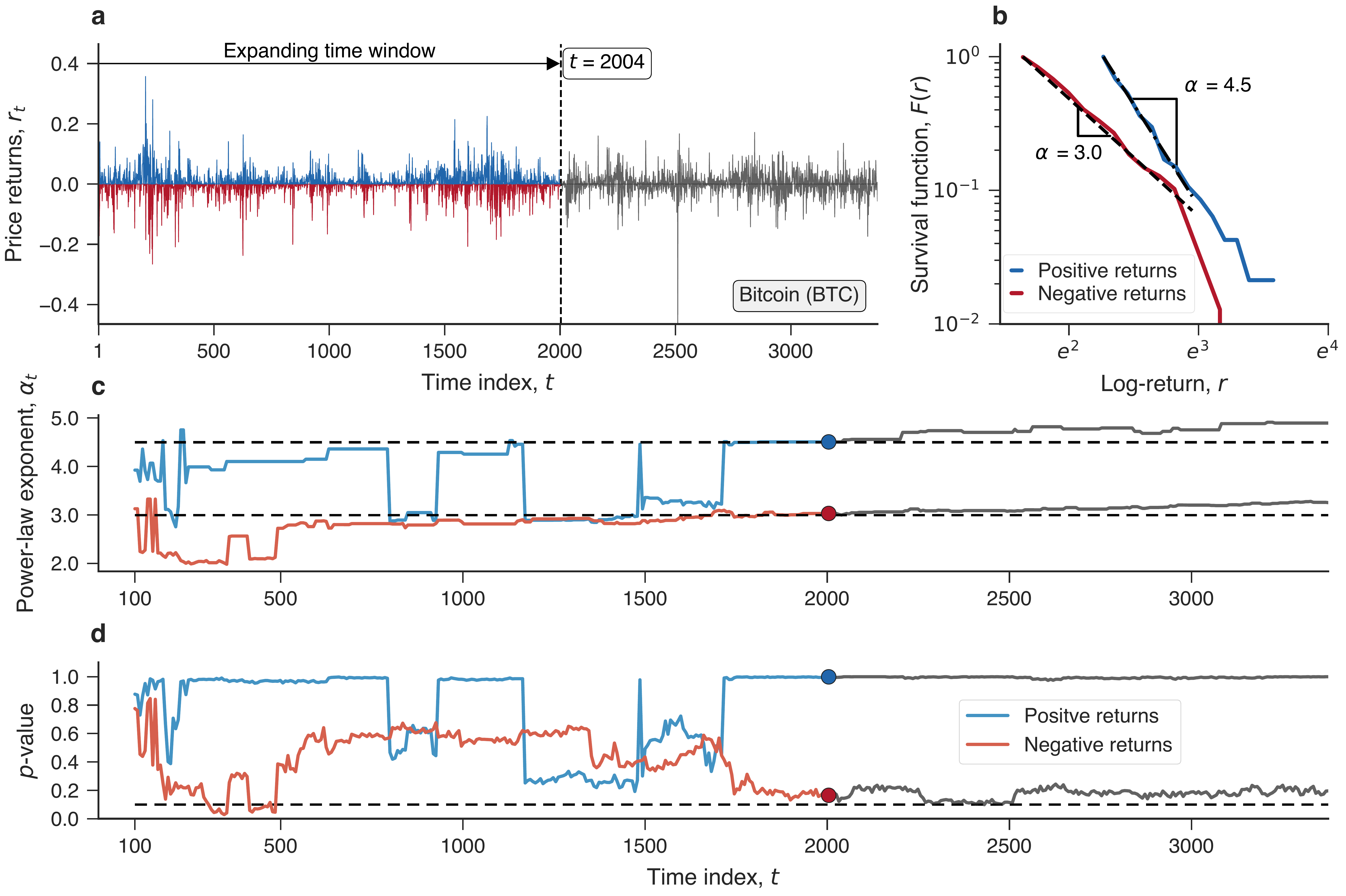}
\caption{Illustration of the approach used to probe patterns in price returns of digital currencies. (\textbf{a}) Bitcoin's time series of daily returns ($r_t$) between 29 April 2013 ($t=1$) and 25 July 2022 ($t=3375$). The black horizontal arrow represents a given position of the expanding time window (at $t=2004$ days) used to sample the return series over the entire history of Bitcoin. This time window expands in weekly steps (seven time series observations), and for each position, we separate the positive (blue) from the negative (red) price returns. The gray line illustrates observations that will be included in future positions of the expanding time window ($t>2004$). (\textbf{b}) Survival functions or the complementary cumulative distributions of positive (blue) and negative (red) price returns within the expanding time window for $t=2004$ days and above the lower bound of the power-law regime estimated from the Clauset-Shalizi-Newman method~\cite{clauset2009powerlaw}. The dashed lines show the adjusted power-law functions, $p(r)\sim r^{-\alpha}$, with $\alpha=4.5$ for positive returns and $\alpha=3.0$ for negative returns. (\textbf{c}) Time series of the power-law exponents $\alpha_t$ for the positive (blue) and negative (red) return distributions obtained by expanding the time window from the hundredth observation ($t=100$) to the latest available price return of Bitcoin. The circular markers represent the values for the window position at $t=2004$ days and the dashed lines indicate the median of the power-law exponents ($\tilde{\alpha}_{\,+}=4.50$ for positive returns and $\tilde{\alpha}_{\,-}=2.99$ for negative returns). (\textbf{d}) Time series of the $p$-values related to the power-law hypothesis of positive (blue) and negative (red) price returns for every position of the expanding time window. The dashed line indicates the threshold ($p=0.1$) above which the power-law hypothesis cannot be rejected. For Bitcoin, the power-law hypothesis is never rejected for positive returns (fraction of rejection $f_r=0$) and rejected in only $4$\% of the expanding time window positions (fraction of rejection $f_r=0.04$).
}
\label{fig:1}
\end{figure*}

Our results are based on daily price time series of 7111 cryptocurrencies that comprise a significant part of all currently available cryptoassets (see Methods for details). From these price series, we have estimated their logarithmic returns
\begin{equation}
    r_t = \ln (x_t / x_{t+1})\,,
\end{equation}
where $x_t$ represents the price of a given cryptocurrency at day $t$. All return time series in our analysis have at least 200 observations (see Supplementary Figure~S1 for the length distribution). Figure~\ref{fig:1}(a) illustrates Bitcoin's series of daily returns. To investigate whether and how returns have changed over the aging and growing processes of all cryptocurrencies, we sample all time series of log-returns using a time window that expands in weekly steps (seven time series observations), starting from the hundredth observation to the latest return observation. In each step, we separate the positive from the negative return values and estimate their power-law behavior using the Clauset-Shalizi-Newman method~\cite{clauset2009powerlaw}. Figure~\ref{fig:1}(a) further illustrates this procedure, where the vertical dashed line represents a given position of the time window ($t=2004$ days), the blue and red lines indicate positive and negative returns, respectively, and the gray lines show the return observations that will be included in the expanding time window in future steps. Moreover, Fig.~\ref{fig:1}(b) shows the corresponding survival functions (or complementary cumulative distributions) for the positive (blue) and negative (red) returns of Bitcoin within the time window highlighted in Fig.~\ref{fig:1}(a). These survival functions correspond to return values above the lower bound of the power-law regime ($r_{\text{min}}$) and dashed lines in Fig.~\ref{fig:1}(b) show the power-law functions adjusted to data, that is,
\begin{equation}
    p(r) \sim r^{-\alpha}\quad(\text{for $r>r_{\text{min}}$})\,,
\end{equation}
with $\alpha=4.5$ for the positive returns and $\alpha=3.0$ for the negative returns in this particular position of the time window ($t=2004$ days).

We have further verified the goodness of the power-law fits using the approach proposed by Clauset \textit{et al.}~\cite{clauset2009powerlaw} (see also Preis \textit{et al.}~\cite{preis2011switching}). As detailed in the Methods section, this approach consists in generating several synthetic samples under the power-law hypothesis, adjusting these simulated samples, and estimating the fraction of times the Kolmogorov-Smirnov distance between the adjusted power-law and the synthetic samples is larger than the value calculated from the empirical data. This fraction defines a $p$-value and allows us to reject or not the power-law hypothesis of the return distributions under a given confidence level. Following Refs.~\cite{clauset2009powerlaw, preis2011switching}, we consider the more conservative 90\% confidence level (instead of the more lenient and commonly used 95\% confidence level), rejecting the power-law hypothesis when $p\text{-value}\leq0.1$. For the particular examples in Fig.~\ref{fig:1}(b), the $p$-values are respectively $1.00$ and $0.17$ for the positive and negative returns, and thus we cannot reject the power-law hypotheses.

After sampling the entire price return series, we obtain time series for the power-law exponents ($\alpha_t$) associated with positive and negative returns as well as the corresponding $p$-values time series for each step $t$ of the expanding time window. These time series allow us to reconstruct the aging process of the return distributions over the entire history of each cryptoasset and probe possible time-dependent patterns. Figures~\ref{fig:1}(c) and \ref{fig:1}(d) show the power-law exponents and $p$-values time series for the case of Bitcoin. The power-law hypothesis is never rejected for positive returns and rarely rejected for negative returns (about $4$\% of times). Moreover, the power-law exponents exhibit large fluctuations at the beginning of the time series and become more stable as Bitcoin matures as a financial asset (a similar tendency as reported by Begu{\v s}i{\'c}~\textit{et al.}~\cite{begusic2018scaling}). The time evolution of these exponents further shows that the asymmetry between positive and negative returns observed in Fig.~\ref{fig:1}(b) is not an incidental feature of a particular moment in Bitcoin's history. Indeed, the power-law exponent for positive returns is almost always larger than the exponent for negative returns, implying that large negative price returns have been more likely to occur than their positive counterparts over nearly the entire history of Bitcoin covered by our data. However, while the difference between positive and negative exponents has approached a constant value, both exponents exhibit an increasing trend, indicating that large price variations are becoming less frequent with the coming-of-age of Bitcoin.

\begin{figure*}[!t]
\centering
\includegraphics[width=1\linewidth]{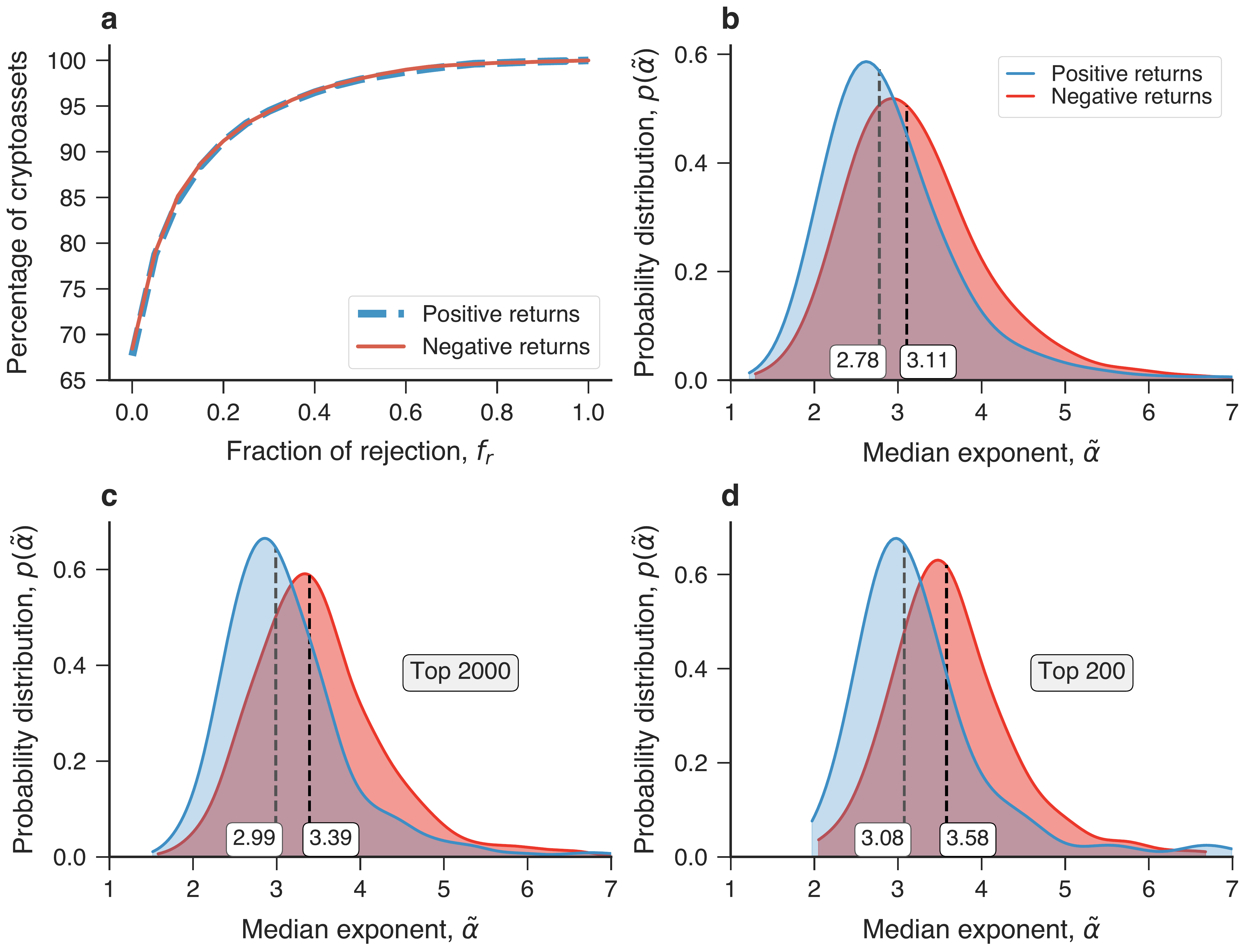}
\caption{Large price movements are power-law distributed over the entire history of most cryptocurrencies with median values typically smaller than those found for traditional assets. (\textbf{a}) Percentage of cryptoassets rejecting the power-law hypothesis for large positive (blue) or negative (red) price returns in at most a given fraction of the weekly positions of the expanding time window ($f_r$) used to sample the return series. Remarkably, $68$\% of all 7111 digital currencies are compatible with the power-law hypothesis over their entire history, and about  $91$\% of them reject the power-law hypothesis in less than 20\% of the positions of the expanding time window ($f_r \leq 0.2$). (\textbf{b}) Probability distributions obtained via kernel density estimation of the median values of the power-law exponents along the history of each digital currency. The blue curve shows the distribution of the median exponents related to positive returns ($\tilde{\alpha}_{\,+}$) and the red curve does the same for negative returns ($\tilde{\alpha}_{\,-}$). The medians of $\tilde{\alpha}_{\,+}$ and $\tilde{\alpha}_{\,-}$ are indicated by vertical dashed lines. Panels (\textbf{c}) and (\textbf{d}) show the distributions of these median exponents when considering the top 2000 and the top 200 cryptocurrencies by market capitalization, respectively. We observe that the distributions of $\tilde{\alpha}_{\,+}$ and $\tilde{\alpha}_{\,-}$ tend to shift toward larger values when considering the largest cryptoassets.
}
\label{fig:2}
\end{figure*}

The previous analysis motivates us to ask whether the entire cryptocurrency market behaves similarly to Bitcoin and what other common patterns digital currencies tend to follow. To start answering this question, we have considered the $p$-values series of all cryptocurrencies to verify if the power-law hypothesis holds in general. Figure~\ref{fig:2}(a) shows the percentage of cryptoassets rejecting the power-law hypothesis in at most a given fraction of the weekly positions of the expanding time window ($f_r$). Remarkably, the hypothesis that large price movements (positive or negative) follow a power-law distribution is never rejected over the entire history of about 70\% of all digital currencies in our dataset. This analysis also shows that only $\approx$$2$\% of cryptocurrencies reject the power-law hypothesis in more than half of the positions of the expanding time window ($f_r\geq0.5$). For instance, considering a 10\% threshold as a criterion ($f_r \leq 0.1$), we find that about 85\% of cryptocurrencies have return distributions adequately modeled by power laws. Increasing this threshold to a more lenient 20\% threshold ($f_r \leq 0.2$), we find large price movements to be power-law distributed for about 91\% of cryptocurrencies. These results thus provide strong evidence that cryptoassets, fairly generally, present large price movements quite well described by power-law distributions. Moreover, this conclusion is robust when starting the expanding window with a greater number of return observations (between 100 and 300 days) and filtering out cryptoassets with missing observations (Supplementary Figures~S2 and S3). Still, it is worth noticing the existence of a few cryptoassets (9 of them) with relatively small market capitalization (ranking below the top 1000) for which the power-law hypothesis is always rejected (Supplementary Table~S1). 

Having verified that large price movements in the cryptocurrency market are generally well-described by power-law distributions, we now focus on the power-law exponents that typically characterize each cryptoasset. To do so, we select all exponent estimates over the entire history of each digital asset for which the power-law hypothesis is not rejected and calculate their median values for both the positive ($\tilde{\alpha}_{\,+}$) and negative ($\tilde{\alpha}_{\,-}$) returns. The dashed lines in Fig.~\ref{fig:1}(c) show these median values for Bitcoin where $\tilde{\alpha}_{\,+}=4.50$ and $\tilde{\alpha}_{\,-}=2.99$. It is worth noticing that the variance of large price movements $\sigma^2$ is finite only for $\alpha>3$, as the integral $\sigma^2 \sim \int_{r_{\text{min}}}^\infty r^2 p(r) dr$ diverges outside this interval. Thus, while the typical variance of large positive returns is finite for Bitcoin, negative returns are at the limit of not having a typical scale and are thus susceptible to much larger variations. Figure~\ref{fig:2}(b) shows the probability distribution for the median power-law exponents of all cryptoassets grouped by large positive and negative returns. We note that the distribution of typical power-law exponents associated with large positive returns is shifted to smaller values when compared with the distribution of exponents related to large negative returns. The medians of these typical exponents are respectively $2.78$ and $3.11$ for positive and negative returns. This result suggests that the asymmetry in large price movements we have observed for Bitcoin is an overall feature of the cryptocurrency market. By calculating the difference between the typical exponents related to positive and negative large returns ($\Delta \alpha = \tilde{\alpha}_{\,+} - \tilde{\alpha}_{\,-}$) for each digital currency, we find that about $2/3$ of cryptocurrencies have $\tilde{\alpha}_{\,+}<\tilde{\alpha}_{\,-}$ (see Supplementary Figure~S4 for the probability distribution of $\Delta \alpha$). Thus, unlike Bitcoin, most cryptocurrencies have been more susceptible to large positive price variations than negative ones. While this asymmetry in the return distributions indicates that extremely large price variations tend to be positive, it does not necessarily imply positive price variations are more common for any threshold in the return values. This happens because the fraction of events in each tail is also related to the lower bound of the power-law regime ($r_{\text{min}}$). However, we have found the distribution of $r_{\text{min}}$ to be similar among the positive and negative returns [Supplementary Figure S5(a)]. The distribution of high percentile scores (such as the 90th percentile) is also shifted to larger values for positive returns [Supplementary Figure S5(b)]. Moreover, this asymmetry in high percentile scores related to positive and negative returns is systematic along the evolution of the power-law exponents [Supplementary Figure S5(c)]. These results thus indicate that there is indeed more probability mass in the positive tails than in the negative ones, a feature that likely reflects the current expansion of the cryptocurrency market as a whole. The distributions in Fig.~\ref{fig:2}(b) also show that large price variations do not have a finite variance for a significant part of cryptoassets, that is, $\tilde{\alpha}_{\,+}\leq3$ for $62$\% of cryptocurrencies and $\tilde{\alpha}_{\,-}\leq3$ for $44$\% of cryptocurrencies. A significant part of the cryptocurrency market is thus prone to price variations with no typical scale. Intriguingly, we further note the existence of a minority group of cryptoassets with $\tilde{\alpha}_{\,+}\leq2$ ($7$\%) or $\tilde{\alpha}_{\,-}\leq2$ ($3$\%). These cryptocurrencies, whose representative members are Counos X (CCXX, rank 216) with $\alpha_{\,-} = 1.96$ and $\alpha_{\,+} = 1.84$ and Chainbing (CBG, rank 236) with $\alpha_{\,+} = 1.87$, are even more susceptible to extreme price variations as one cannot even define the average value $\mu$ for large price returns, as the integral $\mu \sim \int_{r_{\text{min}}}^\infty r p(r) dr$ diverges for $\alpha\leq2$.

We have also replicated the previous analysis when considering cryptocurrencies in the top 2000 and top 200 rankings of market capitalization (as of July 2022). Figures~\ref{fig:2}(c) and~\ref{fig:2}(d) show the probability distribution for the median power-law exponents of these two groups. We observe that these distributions are more localized (particularly for the top 200) than the equivalent distributions for all cryptocurrencies. The fraction of cryptocurrencies with no typical scale for large price returns ($\tilde{\alpha}_{\,+}\leq3$ and $\tilde{\alpha}_{\,-}\leq3$) is significantly lower in these two groups compared to all cryptocurrencies. In the top 2000 cryptocurrencies, $51$\% have $\tilde{\alpha}_{\,+}\leq3$ and $26$\% have $\tilde{\alpha}_{\,-}\leq3$. These fractions are even smaller among the top 200 cryptocurrencies, with only $44$\% and $15$\% not presenting a typical scale for large positive and negative price returns, respectively. We further observe a decrease in the fraction of cryptoassets for which the average value for large price returns is not even finite, as only $2$\% and $1$\% of top 2000 cryptoassets have $\tilde{\alpha}_{\,+}\leq2$ and $\tilde{\alpha}_{\,-}\leq2$. This reduction is more impressive among the top 200 cryptocurrencies as only the cryptoasset Fei USD (FEI, rank 78) has $\tilde{\alpha}_{\,+} = 1.97$ and none is characterized by $\tilde{\alpha}_{\,-}\leq2$. The medians of $\tilde{\alpha}_{\,+}$ and $\tilde{\alpha}_{\,-}$ also increase from $2.78$ and $3.11$ for all cryptocurrencies to $2.98$ and $3.35$ for the top 2000 and to $3.08$ and $3.58$ for the top 200 cryptocurrencies. Conversely, the asymmetry between positive and negative large price returns does not differ much among the three groups, with the condition $\tilde{\alpha}_{\,+}<\tilde{\alpha}_{\,-}$ holding only for a slightly larger fraction of top 2000 ($69.1$\%) and top 200 ($70.6$\%) cryptoassets compared to all cryptocurrencies ($66.4$\%). Moreover, all these patterns are robust when filtering out time series with sampling issues or when considering only cryptoassets that stay compatible with the power-law hypothesis in more than 90\% of the positions of the expanding time window (Supplementary Figures~S6 and~S7).

We also investigate whether the patterns related to the median of the power-law exponents differ among groups of cryptocurrencies with different designs and purposes. To do so, we group digital assets using the 50 most common tags in our dataset (\textit{e.g.} ``bnb-chain'', ``defi'', and ``collectibles-nfts'') and estimate the probability distributions of the median exponents $\tilde{\alpha}_{\,+}$ and $\tilde{\alpha}_{\,-}$ (Supplementary Figures~S8 and~S9). These results show that design and purpose affect the dynamics of large price variations in the cryptocurrency market as the medians of typical exponents range from $2.4$ to $3.7$ among the groups. The lowest values occur for cryptocurrencies tagged as ``doggone-doggerel'' (medians of $\tilde{\alpha}_{\,+}$ and $\tilde{\alpha}_{\,-}$ are $2.38$ and $2.83$), ``memes'' ($2.41$ and $2.87$), and ``stablecoin'' ($2.65$ and $2.79$). Digital currencies belonging to the first two tags overlap a lot and have Dogecoin (DOGE, rank 9) and Shiba Inu (SHIB, rank 13) as the most important representatives. Cryptoassets with these tags usually have humorous characteristics (such as an Internet meme) and several have been considered as a form of pump-and-dump scheme~\cite{kamps2018moon, li2021cryptocurrency, xu2019anatomy}, a type of financial fraud in which false statements artificially inflate asset prices so the scheme operators sell their overvalued cryptoassets. Conversely, cryptoassets tagged as ``stablecoin'' represent a class of cryptocurrencies designed to have a fixed exchange rate to a reference asset (such as a national currency or precious metal)~\cite{eichengreen2019commodity, bullman2019search}. While the price of stablecoins tends to stay around the target values, their price series are also marked by sharp variations, which in turn are responsible for their typically small power-law exponents. This type of cryptoasset has been shown to be prone to failures~\cite{clements2021built, grobys2021stability, briola2022anatomy}, such as the recent examples of TerraUSD (UST) and Tron's USDD (USDD) that lost their pegs to the US Dollar producing large variations in their price series. The asymmetry between positive and negative large returns also emerges when grouping the cryptocurrencies using their tags. All 50 tags have distributions of $\tilde{\alpha}_{\,+}$ shifted to smaller values when compared with the distributions of $\tilde{\alpha}_{\,-}$, with differences between their medians ranging from $-0.74$ (``okex-blockdream-ventures-portfolio'') to $-0.14$ (``stablecoin''). Indeed, only four (`stablecoin'', ``scrypt'', ``fantom-ecosystem'' and ``alameda-research-portfolio'') out of the fifty groupings have both distributions indistinguishable under a two-sample Kolmogorov-Smirnov test ($p$-value $> 0.05$). 

Focusing now on the evolution of the power-law exponents quantified by the time series $\alpha_t$ for positive and negative returns, we ask whether these exponents present particular time trends. For Bitcoin [Fig.~\ref{fig:1}{(c)}], $\alpha_t$ seems to increase with time for both positive and negative returns. At the same time, the results of Fig.~\ref{fig:2} also suggest that market capitalization affects these power-law exponents. To verify these possibilities, we assume the power-law exponents ($\alpha_t$) to be linearly associated with the cryptocurrency's age ($y_t$, measured in years) and the logarithm of market capitalization ($\log c_t$). As detailed in the Methods section, we frame this problem using a hierarchical Bayesian model. This approach assumes that the linear coefficients associated with the effects of age ($A$) and market capitalization ($C$) of each digital currency are drawn from distributions with means $\mu_A$ and $\mu_C$ and standard deviations $\sigma_A$ and $\sigma_C$, which are in turn distributed according to global distributions representing the overall impact of these quantities on the cryptocurrency market. The Bayesian inference process consists of estimating the posterior probability distributions of the linear coefficients for each cryptocurrency as well as the posterior distributions of $\mu_A$, $\mu_C$, $\sigma_A$, and $\sigma_C$, allowing us to simultaneously probe asset-specific tendencies and overall market characteristics. Moreover, we restrict this analysis to the 2140 digital currencies having more than 50 observations of market capitalization concomitantly to the time series of the power-law exponents in order to have enough data points for detecting possible trends.

\begin{figure*}[!ht]
\centering
\includegraphics[width=.86\linewidth]{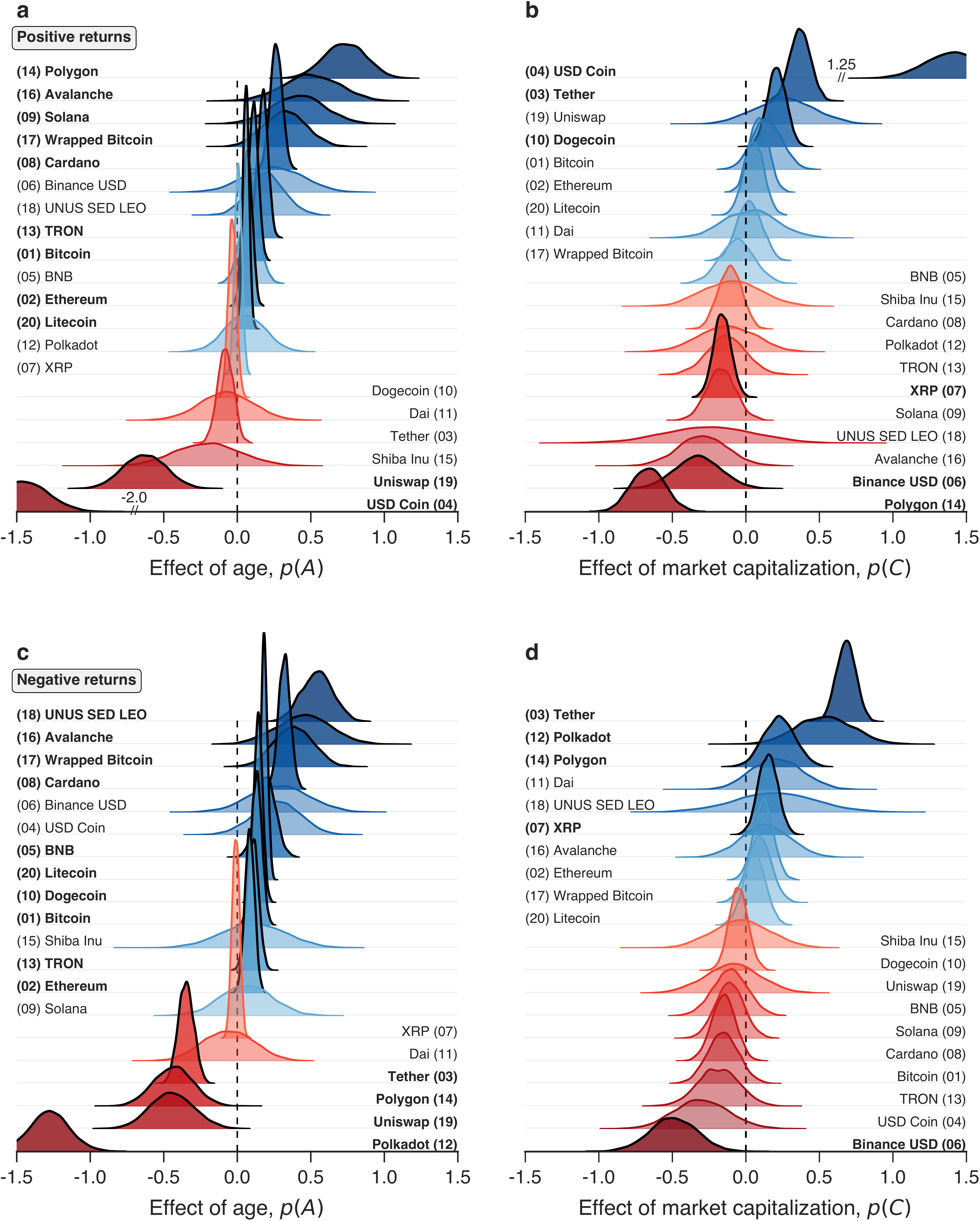}
\caption{Illustration of different effects of age and market capitalization on power-law exponents of cryptocurrencies. (\textbf{a}) Posterior probability distributions of the linear coefficients associated with the effects of age [$p(A)$] and (\textbf{b}) the effects of market capitalization [$p(C)$] on power-law exponents related to large positive returns. Panels (\textbf{c}) and (\textbf{d}) show the analogous distributions for the association with power-law exponents related to large negative returns. In all panels, the different curves show the distributions for each of the top 20 cryptoassets by market capitalization. Cryptocurrencies significantly affected by age or market capitalization are highlighted in boldface, and the numbers between brackets show their positions in the market capitalization rank.
}
\label{fig:3}
\end{figure*}

When considering the overall market characteristics, we find that the $94$\% highest density intervals for $\mu_A$ ([-0.01, 0.06] for positive and [-0.02, 0.03] for negative returns) and $\mu_C$ ([-0.02, 0.03] for positive and [-0.001, 0.04] for negative returns) include the zero (see Supplementary Figure~S10 for their distributions). Thus, there is no evidence of a unique overall pattern for the association between the power-law exponents and age or market capitalization followed by a significant part of the cryptocurrency market. Indeed, the $94$\% highest density intervals for $\sigma_A$ ([0.87, 0.93] for positive and [0.63, 0.70] for negative returns) and $\sigma_C$ ([0.57, 0.61] for positive and [0.49, 0.52] for negative returns) indicate that the cryptocurrency market is highly heterogeneous regarding the evolution of power-law exponents associated with large price variations (see Supplementary Figure~S10 for the distributions of $\sigma_A$ and  $\sigma_C$). Figure~\ref{fig:3} illustrates these heterogeneous behaviors by plotting the posterior probability distributions for the linear coefficients associated with the effects of age ($A$) and market capitalization ($C$) for the top 20 digital assets, where cryptocurrencies which are significantly affected (that is, the $94$\% highest density intervals for $A$ or $C$ do not include the zero) by these quantities are highlighted in boldface. Even this small selection of digital currencies already presents a myriad of patterns. First, we observe that the power-law exponents of a few top 20 cryptocurrencies are neither correlated with age nor market capitalization. That is the case of Shiba Inu (SHIB, rank 13) and Dai (DAI, rank 11) for both positive and negative returns, UNUS SED LEO (LEO, rank 18) and Polkadot (DOT, rank 12) for the positive returns, and USDCoin (USDC, rank 4) and Solana (SOL, rank 9) for negative returns. There are also cryptocurrencies with exponents positively or negatively correlated only with market capitalization. Examples include Tether (USDT, rank 3) and Dogecoin (DOGE, rank 10), for which the power-law exponents associated with positive returns increase with market capitalization, and Binance USD (BUSD, rank 6), for which power-law exponents associated with positive and negative returns decrease with market capitalization. We also observe cryptocurrencies for which age and market capitalization simultaneously affect the power-law exponents. Polygon (MATIC, rank 14) is an example where the power-law exponents associated with positive returns tend to increase with age and decrease with market capitalization. Finally, there are also cryptocurrencies with power-law exponents only associated with age. That is the case of Bitcoin (BTC, rank 1), Ethereum (ETH, rank 2), and Cardano (ADA, rank 8), for which the power-law exponents related to positive and negative returns increase with age, but also the case of Uniswap (UNI, rank 19), for which the exponents decrease with age.

\begin{figure*}[!ht]
\centering
\includegraphics[width=0.96\linewidth]{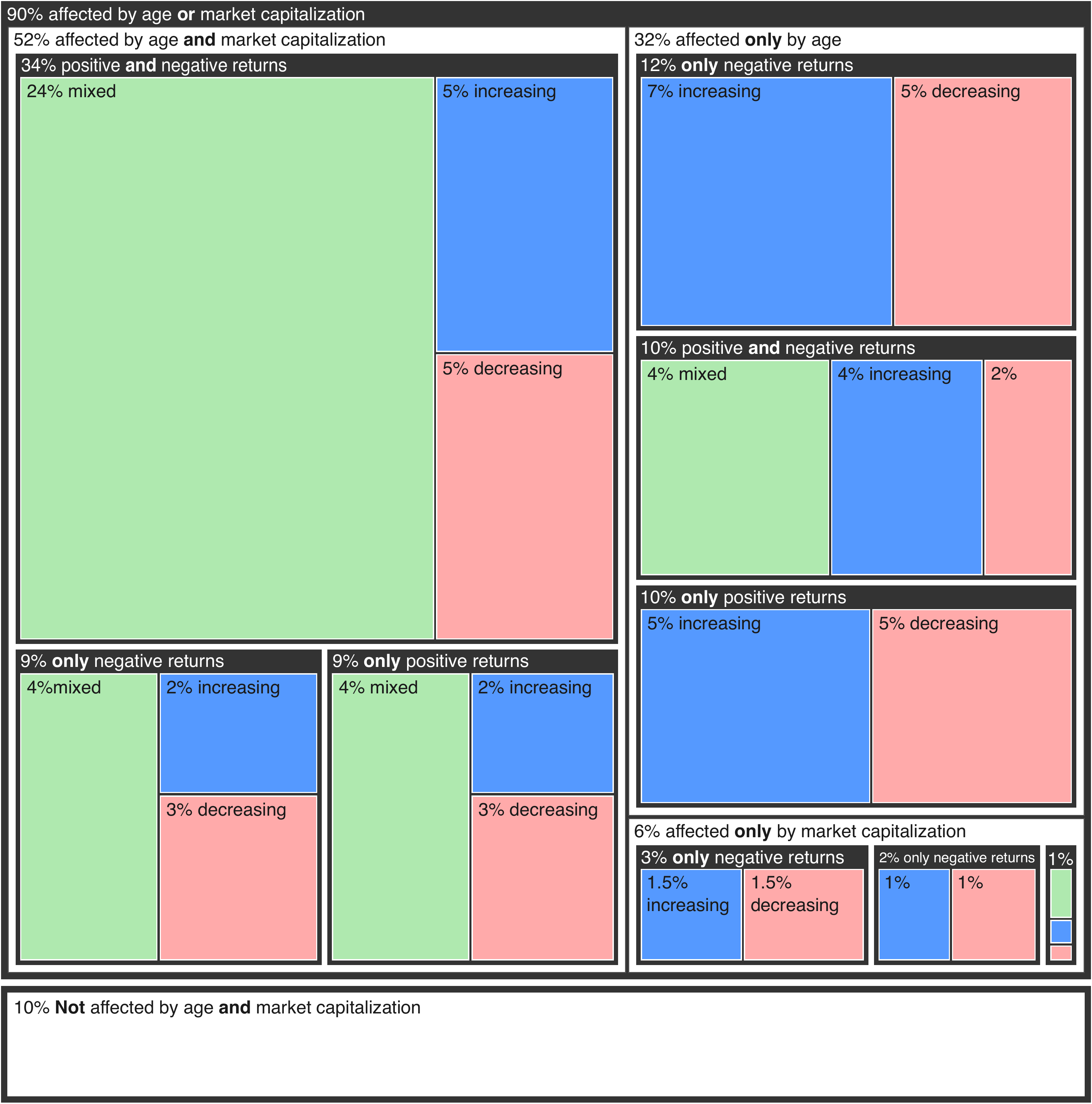}
\caption{Summary of the effects of age and market capitalization on power-law exponents of the cryptocurrency market. Hierarchical visualization or a tree map of the possible effects of age and market capitalization on the power-law exponents. The first level (two outermost rectangles) separates cryptocurrencies that are affected by age or market capitalization (90\%) from those unaffected by any of these quantities (10\%). Cryptocurrencies affected by age or market capitalization are classified as those simultaneously affected by both quantities (52\%), those affected only by age (32\%), and those affected only by market capitalization (6\%). Each of the previous three levels is further classified regarding whether both positive and negative returns are simultaneously affected or whether the effect involves only positive or only negative returns. Finally, the former levels are classified regarding whether the power-law exponents increase, decrease or have a mixed trend with the predictive variables. Overall, 36\% of the associations are classified as mixed trends (green rectangles), 28\% are increasing trends (blue rectangles), and 26\% are decreasing trends (red rectangles). 
}
\label{fig:4}
\end{figure*}

Figure~\ref{fig:4} systematically extends the observations made for the top 20 cryptoassets to all 2140 digital currencies for which we have modeled the changes in the power-law exponents as a function of age and market capitalization. First, we note that only 10\% of cryptocurrencies have power-law exponents not significantly affected by age and market capitalization. The vast majority (90\%) displays some relationship with these quantities. However, these associations are as varied as the ones we have observed for the top 20 cryptoassets. About 52\% of cryptocurrencies have power-law exponents simultaneously affected by age and market capitalization. In this group, these quantities simultaneously impact the exponents related to positive and negative returns of 34\% of cryptoassets, whereas the remainder is affected only in the positive tail (9\%) or only in the negative tail (9\%). Moving back in the hierarchy, we find that the power-law exponents of 32\% of cryptocurrencies are affected only by age while a much minor fraction  (6\%) is affected only by market capitalization. Within the group only affected by age, we observe that the effects are slightly more frequent only on the exponents related to negative returns (12\%), compared to cases where effects are restricted only to positive returns (10\%) or simultaneously affect both tails (10\%). Finally, within the minor group only affected by market capitalization, we note that associations more frequently involve only exponents related to negative returns (3\%) compared to the other two cases (2\% only positive returns and 1\% for both positive and negative returns).

Beyond the previous discussion about whether positive or negative returns are simultaneously or individually affected by age and market capitalization, we have also categorized the direction of the trend imposed by these two quantities on the power-law exponents. Blue rectangles in Fig.~\ref{fig:4} represent the fraction of relationships for which increasing age or market capitalization (or both) is associated with a raise in the power-law exponents. About 28\% of all cryptocurrencies exhibit this pattern in which large price variations are expected to occur less frequently as they grow and age. Conversely, the red rectangles in Fig.~\ref{fig:4} depict the fraction of relationships for which increasing age or market capitalization (or both) is associated with a reduction in the power-law exponents. This case comprises about 25\% of all cryptocurrencies for which large price variations are likely to become more frequent as they grow in market capitalization and age. Still, the majority of associations represented by green rectangles refer to the case where the effects of age and market capitalization point in different directions (\textit{e.g.} exponents increasing with age while decreasing with market capitalization). About 36\% of cryptocurrencies fit this condition which in turn contributes to consolidating the cumbersome hierarchical structure of patterns displayed by cryptocurrencies regarding the dynamics of large price variations. This complex picture is not much different when considering only cryptocurrencies in the top 200 market capitalization rank (Supplementary Figure~S11). However, we do observe an increased prevalence of patterns characterized by exponents that rise with age and market capitalization (37\%), suggesting that large price variations are becoming less frequent among the top 200 cryptocurrencies than in the overall market. 

\section*{Discussion}

We have studied the distributions of large price variations of a significant part of the digital assets that currently comprise the entirety of the cryptocurrency market. Unlike previous work, we have estimated these distributions for entire historical price records of each digital currency, and we have identified the patterns under which the return distributions change as cryptoassets age and grow in market capitalization. Similarly to conventional financial assets~\cite{gopikrishnan1998inverse, mantegna1999introduction, gabaix2003theory}, our findings show that the return distributions of the vast majority of cryptoassets have tails that are described well by power-law functions along their entire history. The typical power-law exponents of cryptocurrencies ($\alpha\sim3$) are, however, significantly smaller than those reported for conventional assets ($\alpha\sim4$)~\cite{gopikrishnan1998inverse, mantegna1999introduction, gabaix2003theory}. This feature corroborates the widespread belief that cryptoassets are indeed considerably more risky for investments than stocks or other more traditional financial assets. Indeed, we have found that about half of the cryptocurrencies in our analysis do not have a characteristic scale for price variations, and are thus prone to much higher price variations than those typically observed in stock markets. On the upside, we have also identified an asymmetry in the power-law exponents for positive and negative returns in about $2/3$ of all considered cryptocurrencies, such that these exponents are smaller for positive than they are for negative returns. This means that sizable positive price variations have generally been more likely to occur than equally sizable negative price variations, which in turn may also reflect the recent overall expansion of the cryptocurrency market.

Using a hierarchical Bayesian linear model, we have also simultaneously investigated the overall market characteristics and asset-specific tendencies regarding the effects of age and market capitalization on the power-law exponents. We have found that the cryptocurrency market is highly heterogeneous regarding the trends exhibited by each cryptocurrency; however, only a small fraction of cryptocurrencies (10\%) have power-law exponents neither correlated with age nor market capitalization. These associations have been mostly ignored by the current literature and are probably related to the still-early developmental stage of the cryptocurrency market as a whole. Overall, 36\% of cryptocurrencies present trends that do not systematically contribute to increasing or decreasing their power-law exponents as they age and grow in market capitalization. On the other hand, for 26\% of cryptocurrencies, aging and growing market capitalization are both associated with a reduction in their power-law exponents, thus contributing to the rise in the frequency of large price variations in their dynamics. Only about 28\% of cryptocurrencies present trends in which the power-law exponents increase with age and market capitalization, favoring thus large price variations to become less likely. These results somehow juxtapose with findings about the increasing informational efficiency of the cryptocurrency market~\cite{sigaki2019clustering}. In fact, if on the one hand the cryptocurrency market is becoming more informationally efficient, then on the other our findings indicate that there is no clear trend toward decreasing the risks of sizable variations in the prices of most considered cryptoassets. In other words, risk and efficiency thus appear to be moving towards different directions in the cryptocurrency market.

To conclude, we hope that our findings will contribute significantly to the better understanding of the dynamics of large price variations in the cryptocurrency market as a whole, and not just for a small subset of selected digital assets, which is especially relevant due to the diminishing concentration of market capitalization among the top digital currencies, and also because of the considerable impact these new assets may have in our increasingly digital economy.

\section*{Methods}~\label{sec:methods}

\subsection*{Data}~\label{subsec:data}

Our results are based on time series of the daily closing prices (in USD) for all cryptoassets listed on CoinMarketCap (\url{coinmarketcap.com}) as of 25 July 2022 [see Supplementary Figure~S1(a) for a visualization of the increasing number cryptoassets listed on CoinMarketCap since 2013]. These time series were automatically gathered using the \texttt{cryptoCMD} Python package~\cite{gupta2022cryptocmd} and other information such as the tags associated with each cryptoasset were obtained via the CoinMarketCap API~\cite{coinmarketcap-api}. In addition, we have also obtained the daily market capitalization time series (in USD) from all cryptoassets which had this information available at the time. Earliest records available from CoinMarketCap date from 29 April 2013 and the latest records used in our analysis correspond to 25 July 2022. Out of 9943 cryptocurrencies, we have restricted our analysis to the 7111 with at least 200 price-return observations. The median length of these time series is 446 observations [see the distribution of series length in Supplementary Figure~S1(b)]. 

\subsection*{Estimating power-law exponents}~\label{subsec:fit}

We have estimated the power-law behavior of the return distributions by applying the Clauset-Shalizi-Newman method~\cite{clauset2009powerlaw} to the return time series $r_t$. In particular, we have sampled each of these time series using an expanding time window that starts at the hundredth observation and grows in weekly steps (seven data points each step). For each position of the expanding time window, we have separated the positive returns from the negative ones and applied the Clauset-Shalizi-Newman method~\cite{clauset2009powerlaw} to each set. This approach consists of obtaining the maximum likelihood estimate for the power-law exponent, $\alpha = 1 + n / \left(\sum_{t=1}^n \ln {r_t}/{r_{\text{min}}}\right),$ where $r_{\text{min}}$ is the lower bound of the power-law regime and $n$ is the number of (positive or negative) return observations in the power-law regime for a given position of the expanding time window. The value $r_{\text{min}}$ is estimated from data by minimizing the Kolmogorov-Smirnov statistic between the empirical distribution and the power-law model. The Clauset-Shalizi-Newman method~\cite{clauset2009powerlaw} yields an unbiased and consistent estimator~\cite{bhattacharya2020consistency}, in a sense that as the sample increases indefinitely, the estimated power-law exponent converges in distribution to the actual value. Moreover, we have used the implementation available on the \texttt{powerlaw} Python package~\cite{alstott2014powerlaw}.

In addition to obtaining the power-law exponents, we have also verified the adequacy of the power-law hypothesis using the procedure originally proposed by Clauset \textit{et al.}~\cite{clauset2009powerlaw} as adapted by Preis \textit{et al.}~\cite{preis2011switching}. This procedure consists of generating synthetic samples under the power-law hypothesis with the same properties of the empirical data under analysis (that is, same length and parameters $\alpha$ and $r_{\text{min}}$), adjusting the simulated data with the power-law model via the Clauset-Shalizi-Newman method, and calculating the Kolmogorov-Smirnov statistic ($\kappa_{\text{syn}}$) between the distributions obtained from the simulated samples and the adjusted power-law model. Next, the values of $\kappa_{\text{syn}}$ are compared to the Kolmogorov-Smirnov statistic calculated between empirical data and the power-law model ($\kappa$). Finally, a $p$-value is defined by calculating the fraction of times for which $\kappa_{\text{syn}}>\kappa$. We have used one thousand synthetic samples for each position of the expanding time window and the more conservative 90\% confidence level (instead of the more lenient and commonly used 95\% confidence level), such that the power-law hypothesis is rejected whenever $p$-value $\leq0.1$.

\subsection*{Modelling the effects of age and market capitalization on the power-law exponents}~\label{subsec:hierarchical}

We have estimated the effects of age and market capitalization on the power-law exponents associated with positive or negative returns of a given cryptocurrency using the linear model
\begin{equation}\label{eq:cap_time_model}
    \alpha_t \sim \mathcal{N}(K + C\, \log c_t + A\, y_t, \epsilon)\,,
\end{equation}
where $\alpha_{t}$ represents the power-law exponent, $\log c_t$ is the logarithm of the market capitalization, and $y_t$ is the age (in years) of the cryptocurrency at $t$-th observation. Moreover, $K$ is the intercept of the association, while $C$ and $A$ are linear coefficients quantifying the effects of market capitalization and age, respectively. Finally, $\mathcal{N}(\mu, \sigma)$ stands for the normal distribution with mean $\mu$ and standard deviation $\sigma$, such that the parameter $\epsilon$ accounts for the unobserved determinants in the dynamics of the power-law exponents. We have framed this problem using the hierarchical Bayesian approach such that each power-law exponent $\alpha_t$ is nested within a cryptocurrency with model parameters considered as random variables normally distributed with parameters that are also random variables. Mathematically, for each cryptocurrency, we have
\begin{equation}\label{eq:prior_coeffs}
\begin{split}
    K \sim \mathcal{N}(\mu_K, \sigma_K)\,, \quad
    C \sim \mathcal{N}(\mu_C, \sigma_C)\,, \quad
    A \sim \mathcal{N}(\mu_A, \sigma_A)\,,
\end{split}
\end{equation}
where $\mu_K$, $\sigma_K$, $\mu_C$, $\sigma_C$, $\mu_A$, and $\sigma_A$ are hyperparameters. These hyperparameters are assumed to be distributed according to distributions that quantify the overall impact of age and market capitalization on the cryptocurrency market as a whole. 

We have performed this Bayesian regression for exponents related to positive and negative returns separately, and used noninformative prior and hyperprior distributions in order not to bias the posterior estimation~\cite{gelman2007data}. Specifically, we have considered
\begin{equation}\label{eq:hyperpriors}
\begin{split}
    \mu_K^{}    &\sim \mathcal{N}(0, 10^5)\,, \quad
    \sigma_K^{}  \sim {\rm Inv{-}\Gamma}(1, 1)\,,\\
    \mu_C^{}    &\sim \mathcal{N}(0, 10^5)\,, \quad
    \sigma_C^{}  \sim {\rm Inv{-}\Gamma}(1, 1)\,,\\
    \mu_A^{}    &\sim \mathcal{N}(0, 10^5)\,, \quad
    \sigma_A^{}  \sim {\rm Inv{-}\Gamma}(1, 1)\,,\\
\end{split}
\end{equation}
and  $\epsilon \sim \mathcal{U}(0,10^2)\,,$ where $\mathcal{U}(a,b)$ stands for the uniform distribution in the interval $[a,b]$ and ${\text{Inv}{-}\Gamma}(\theta, \gamma)$ represents the inverse gamma distribution with shape and scale parameters $\theta$ and $\gamma$, respectively. For the numerical implementation, we have relied on the PyMC~\cite{salvatier2016probabilistic} Python package and sampled the posterior distributions via the gradient-based Hamiltonian Monte Carlo no-U-Turn-sampler method. We have run four parallel chains with 2500 iterations each (1000 burn-in samples) to allow good mixing and estimated the Gelman-Rubin convergence statistic (R-hat) to ensure the convergence of the sampling approach (R-hat was always close to one).

In addition, we have also verified that models describing the power-law exponents as a function of only age ($C\to0$ in Eq.~\ref{eq:cap_time_model}) or only market capitalization ($A\to0$ in Eq.~\ref{eq:cap_time_model}) yield significantly worse descriptions of our data as quantified by the Widely Applicable Information Criterion (WAIC) and the Pareto Smoothed Importance Sampling Leave-One-Out cross-validation (PSIS-LOO)~\cite{vehtari2016practical} (see Supplementary Table~S2).

\section*{Acknowledgements}

The authors acknowledge the support of the Coordena\c{c}\~ao de Aperfei\c{c}oamento de Pessoal de N\'ivel Superior (CAPES), the Conselho Nacional de Desenvolvimento Cient\'ifico e Tecnol\'ogico (CNPq -- Grant 303533/2021-8), and the Slovenian Research Agency (Grants J1-2457 and P1-0403). The authors are also grateful to Andre Seiji Sunahara for the assistance with the implementation of Bayesian models.

\section*{Author contributions statement}

A.A.B.P., M.P., and H.V.R. designed research, performed research, analyzed data, and wrote the paper.
 
\section*{Data availability}

Data and code necessary to reproduce all results presented in this manuscript are available at \url{gitlab.com/arthurpessa/crypto-returns}.

\bibliography{references}

\clearpage
\includepdf[pages=1-13,pagecommand={\thispagestyle{empty}}]{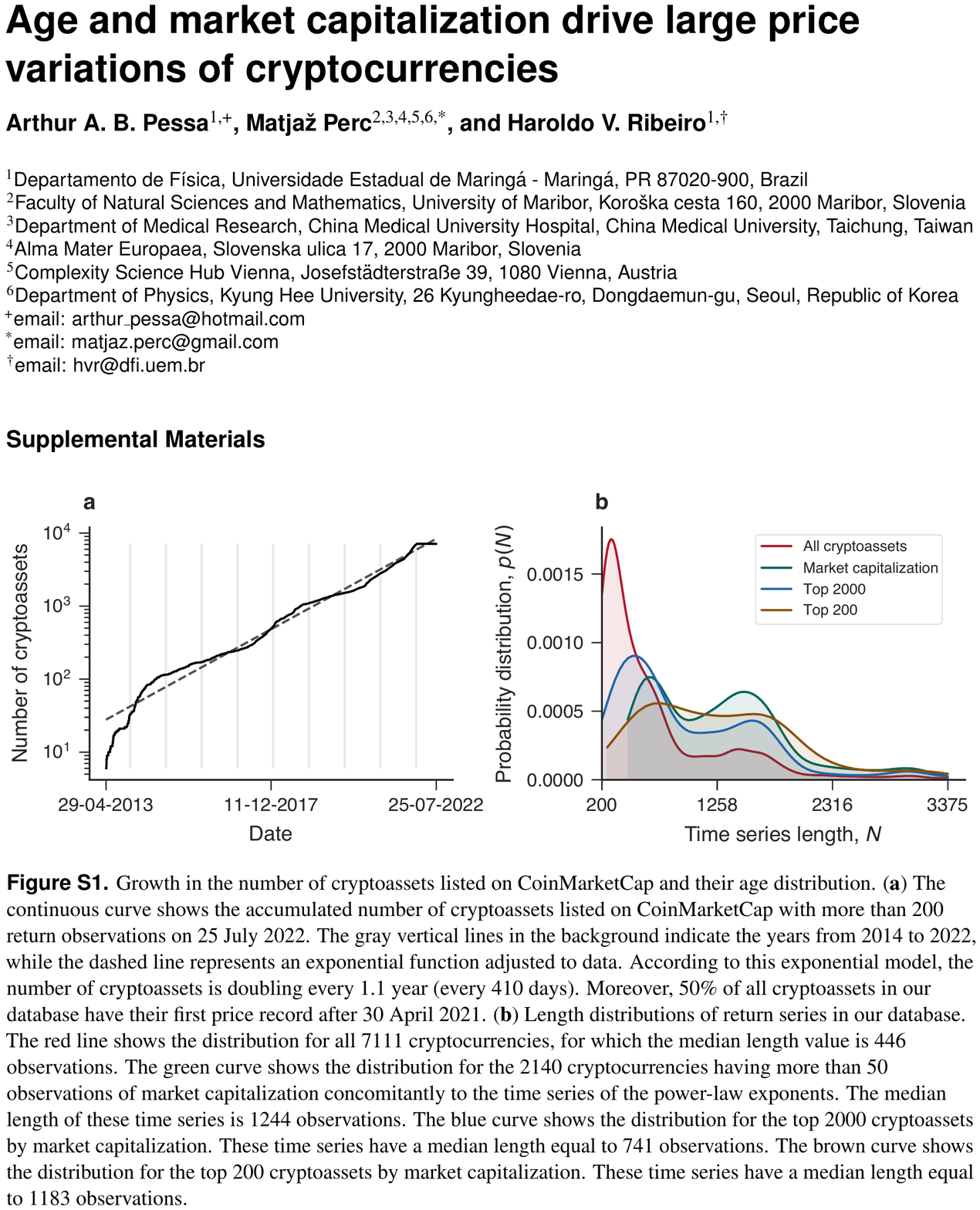}

\end{document}